# Impurity band conduction in group-IV ferromagnetic semiconductor $Ge_{1-x}Fe_x$ with nanoscale fluctuations in Fe concentration

Yoshisuke Ban,[1] Yuki K. Wakabayashi,[1] Ryosho Nakane,[1] and Masaaki Tanaka[1,2]

[1]*Department of Electrical Engineering and Information Systems, The University of Tokyo, 7-3-1 Hongo, Bunkyo-ku, Tokyo 113-8656, Japan*

[2]*Center for Spintronics Research Network, Graduate School of Engineering, The University of Tokyo, 7-3-1 Hongo, Bunkyo-ku, Tokyo 113-8656, Japan*

We study the carrier transport and magnetic properties of group-IV-based ferromagnetic semiconductor $Ge_{1-x}Fe_x$ thin films (Fe concentration $x$ = 2.3 – 14 %) with and without boron (B) doping, by measuring their transport characteristics; the temperature dependence of resistivity, hole concentration, mobility, and the relation between the anomalous Hall conductivity versus conductivity. At relatively low $x$ (= 2.3 %), the transport in the undoped $Ge_{1-x}Fe_x$ film is dominated by hole hopping between Fe-rich hopping sites in the Fe impurity band, whereas that in the B-doped $Ge_{1-x}Fe_x$ film is dominated by the holes in the valence band in the degenerated Fe-poor regions. As $x$ increases ($x$ = 2.3 – 14 %), the transport in the both undoped and B-doped $Ge_{1-x}Fe_x$ films is dominated by hole hopping between the Fe-rich hopping sites of the impurity band. The magnetic properties of the $Ge_{1-x}Fe_x$ films are studied by various methods including magnetic circular dichroism, magnetization and anomalous Hall resistance, and are not influenced by B-doping. We show band profile models of both undoped and B-doped $Ge_{1-x}Fe_x$ films, which can explain the transport and the magnetic properties of the $Ge_{1-x}Fe_x$ films.

**I. INTRODUCTION**

In the past two decades, numerous studies on ferromagnetic semiconductors (FMSs) have been extensively carried out, since FMSs have many potential advantages over metallic ferromagnets, such as their material compatibility with semiconductor devices and possibility of utilizing the band engineering of semiconductors. Therefore, FMSs are expected to be used for novel semiconductor-based electronic devices utilizing spin degrees of freedom as well as charge transport of carriers. Among them group-IV based FMSs are very attractive, because they are excellently compatible with the existing complementary metal-oxide-semiconductor (CMOS) technology. However, preparing a single-crystalline group-IV FMS film of high quality seems to be difficult, since there are far less studies on group-IV FMSs than those on III-V-based FMSs, such as GaMnAs and InMnAs. For example, $Ge_{1-x}Mn_x$ films[1] are not structurally uniform, since intermetallic Mn-Ge compounds are easily formed.[2–7] On the other hand, single-crystalline $Ge_{1-x}Fe_x$ films of diamond-type crystal structure developed by our group possibly have the above-mentioned potential advantages of FMS.[8–14] Here, we briefly review the previous studies on this material, and show unclarified subjects, and then describe the purpose of this study.

Group-IV FMS $Ge_{1-x}Fe_x$ films are grown on Si(001) and Ge(001) substrates by low-temperature molecular beam epitaxy (MBE).[8–14] The crystal structure of $Ge_{1-x}Fe_x$ films is of diamond-type without intermetallic Fe-Ge compounds and other second-phase precipitates, which were revealed by reflection high-energy electron diffraction (RHEED), X-ray diffraction (XRD), cross-sectional high-resolution transmission electron microscopy (TEM) with spatially-resolved transmission-electron diffraction (TED), and energy-dispersive X-ray spectroscopy (EDX).[8] At the same time, the Fe concentration was found to spatially fluctuate: For example, the local Fe concentrations of Fe-rich and Fe-poor regions in a $Ge_{1-x}Fe_x$ film with the average Fe concentration $x = 9.5$ % was ~4 and ~12 %, respectively.[8] Further structural characterizations using channeling Rutherford backscattering (c-RBS) and channeling particle-induced X-ray emission (c-PIXE) measurements[9] revealed that about 80% of the Fe atoms are located at the substitutional sites of the diamond structure in $Ge_{1-x}Fe_x$ ($x = 6.5$ %) films. All these studies verify that $Ge_{1-x}Fe_x$ films have a single-crystalline diamond-type crystal structure.

Single-phase ferromagnetic properties of $Ge_{1-x}Fe_x$ films were confirmed by magnetic circular dichroism (MCD) spectra with various magnetic fields,[10] hysteresis of the anomalous Hall resistance,[11] and temperature dependence of the magnetization (*M-T* curves) and magnetic field dependence of the magnetization (*M-H* curves) measured by a superconducting quantum interference device (SQUID) magnetometer.[12] The Curie temperature $T_C$ was investigated in conjunction with the mechanism of the ferromagnetic order. It was found that $T_C$ increases with increasing $x$, and the highest $T_C = 210$ K was demonstrated in a $Ge_{1-x}Fe_x$ film with $x = 10.5$ % which was annealed at 500°C after the MBE growth.[12] A notable suggestion is that $T_C$ is determined not only by the average Fe concentration $x$ that is estimated by the Fe flux during the growth, but also by the local fluctuation of the Fe concentration in the $Ge_{1-x}Fe_x$ films;[12] $T_C$ increases with increasing the fluctuation. It was recently shown that nanoscale local ferromagnetic regions, which are formed through ferromagnetic exchange interactions in the high-Fe-content regions of the $Ge_{1-x}Fe_x$ films, exist even at room temperature, well above the $T_C$ of 20 – 100 K.[13] It was also found that with decreasing temperature, the local ferromagnetic regions are expanded, followed by a transition of the entire film into a ferromagnetic state at $T_C$.[13] Since the Fe-rich region probably has a significant amount of holes, the double exchange interaction between the Fe atoms most likely induces the ferromagnetic order in that region. However, it was not clarified whether the ferromagnetic order is also induced by the transport carriers (holes) between the Fe-rich regions via the Fe-poor region (lower $x$ region) or not.

The electronic structure of a $Ge_{1-x}Fe_x$ ($x$ = 6.5 %) film was recently studied by soft X-ray angle-resolved photoemission spectroscopy (SX-ARPES) measurements.[15] Although the band structure of the $Ge_{1-x}Fe_x$ film is basically the same as that of Ge, an additional Fe 3d energy band in the band gap, which is located at ~0.35 eV above the valence band top, was also detected. Furthermore, the Fermi level $E_F$ was located at this additional energy band, which comes from the impurity levels of the Fe 3d states. Another study using X-ray magnetic circular dichroism (XMCD)[13] indicates that the substitutional Fe atoms in the $Ge_{1-x}Fe_x$ films ($x$ = 6.5 %) are divalent ($Fe^{2+}$). Thus, these two results are consistent with the model mentioned in the previous paragraph[13] that a large number of holes are present in the Fe-rich regions and induce the ferromagnetic order in these regions. However, the XMCD[13] and SX-ARPES[15] signals did not give direct evidence of the model, since they were composed of both signals from nano-scale Fe-rich and Fe-poor regions.

In our previous work, the electrical transport properties of $Ge_{1-x}Fe_x$ films were also studied. $Ge_{1-x}Fe_x$ films have positive Hall resistances with hysteresis.[14] Thus, mobile carriers in $Ge_{1-x}Fe_x$ films are holes (p-type). On the other hand, recently, presence of spin-polarized carriers in a $Ge_{1-x}Fe_x$ ($x$ = 6.5 %) film has been demonstrated by observing tunnel magnetoresistance (TMR) in an epitaxial Fe(10 nm)/MgO(3 nm)/$Ge_{1-x}Fe_x$ ($x$ = 6.5 %)/Ge(001) junction at 3.5 K.[16] This indicates that $Ge_{1-x}Fe_x$ can be used for spin transport devices. However, there are still unsolved questions in the hole transport of $Ge_{1-x}Fe_x$: Especially, how do the Fe concentration $x$ and its local fluctuation affect the transport? What are the relations between the structural, transport, and magnetic properties? What are the band profiles of $Ge_{1-x}Fe_x$ taking into account the local fluctuation of the Fe concentration?

To answer these questions, in this paper, we investigate the systematic and detailed electrical transport properties of $Ge_{1-x}Fe_x$ without and with boron (B) doping. Here, B works as an acceptor[14] and thus we can increase the hole concentration. Since transport measurements can detect signals arising from small-scale regions, the mechanism of ferromagnetic ordering[13,15] can be examined microscopically. To obtain the precise transport properties, the $Ge_{1-x}Fe_x$ films were grown on silicon-on-insulator (SOI) substrates with a very thin (~5 nm) Si top layer as shown in Fig. 1, by which we can prevent parallel conduction through the substrate. More specifically, one of our purposes is to investigate whether the holes conducting between the Fe-rich regions play an important role in the ferromagnetic order of the entire $Ge_{1-x}Fe_x$ films or not, by comparing the properties of undoped and B-doped $Ge_{1-x}Fe_x$ films with various $x$. Since B doping causes a higher hole density in the Fe-poor regions, it is expected that $T_C$ becomes higher if the holes in the Fe-poor regions induce ferromagnetic coupling between the ferromagnetic Fe-rich regions. Another purpose is to obtain the band profile of the $Ge_{1-x}Fe_x$ films taking into account the spatial fluctuation of the Fe concentration. As described previously, the band structure revealed by SX-ARPES[15] originates from both the Fe-rich and Fe-poor regions. Since the transport properties are expected to change depending on $x$ and undoped/B-doped $Ge_{1-x}Fe_x$ films, we can expect that the properties of both the Fe-rich and Fe-poor regions are revealed by comparing the results of $Ge_{1-x}Fe_x$ films with various material parameters.

In the following sections, we present the crystal growth, magnetic properties (Section II), transport properties involving hopping (Section III A–D), the band profiles of the $Ge_{1-x}Fe_x$ films with $x = 2.3 – 14.0$ % with and without B doping (Section III E and F), and then discuss the relation between the hole transport and the magnetic properties (Section IV). We also describe other transport properties; the anomalous Hall conductivity, conductivity, and temperature dependence of the Hall mobility of the $Ge_{1-x}Fe_x$ films ($x = 2.3 – 14.0$ %) with and without B doping (Section V). Finally, concluding remarks are stated (Section VI).

## II. CRYSTAL GROWTH AND MAGNETIC PROPERTIES

As shown in Fig. 1, we grew single layer $Ge_{1-x}Fe_x$ ($x = 2.3 – 14.0$ %, 100 nm) samples with and without boron (B) doping ($y = 4.4 \times 10^{19}$ cm$^{-3}$, 0) by MBE on SOI substrates composed of undoped Si (~5nm) /SiO$_2$ (50 nm)/Si (001) substrates. The detailed growth process is as follows: First, a (001)-oriented SOI substrate was thermally oxidized and etched with HF to form a 5-nm-thick Si top layer. After thermal cleaning of the SOI substrate at 760°C in our MBE chamber, the substrate temperature was cooled to 200°C. Then, a 100-nm-thick $Ge_{1-x}Fe_x$ ($x = 2.3 – 14.0$%) film was grown at 200°C epitaxially on the SOI (001) substrate by supplying Ge and Fe fluxes by MBE. When B-doped $Ge_{1-x}Fe_x$ ($Ge_{1-x}Fe_x$:B) films were grown, a B flux was also supplied during the growth, where the B doping concentration $y$ was fixed. A B-doped 100-nm-thick Ge (Ge:B) film was also grown on a SOI substrate at 200°C as a reference sample. The parameters $x$ and $y$ of all the $Ge_{1-x}Fe_x$ films are listed in Table I, where $x$ and $y$ were estimated by Rutherford backscattering spectrometry (RBS) and secondary ion mass spectrometry (SIMS), respectively. The magnetic properties of the as-grown samples were characterized by reflection magnetic circular dichroism (MCD) spectra, MCD *vs*. magnetic field (MCD-$H$), and magnetization *vs*. magnetic field ($M$-$H$) measured by a superconducting interference device magnetometer (SQUID). These measurements were performed in the temperature range of 5 – 100 K, and the magnetic field was applied perpendicular to the film plane.

To measure the anomalous Hall resistance (AHR) and resistivity $\rho$, some of the samples were patterned into Hall-bar shaped devices (length: 200 μm, width: 50 μm) by photolithography and wet etching. These transport measurements were performed with a self-made cryostat system with a 1 T electromagnet and Spectromag 4000 (Oxford Instruments) with a 7 T superconducting magnet.

The Curie temperature $T_C$ was estimated by the Arrott plots of MCD-$H$ curves measured at various temperatures (5 – 120 K) as shown in Table I (the definition of $T_C$ in this study is stated in Section I of Supplementary Material (SM)). To see whether the $Ge_{1-x}Fe_x$:B ($x$ = 10.5, 14.0 %) films have a single ferromagnetic phase or not, $M$-$H$ curves, AHR, and MCD-$H$ of the $Ge_{1-x}Fe_x$:B ($x$ = 10.5, 14.0 %) films were measured and compared, as shown in Figs. 2 (a)(b)(c) and (d), where each signal was normalized by each maximum value. Since the normalized signals for the same sample were identical, the magnetic properties in these $Ge_{1-x}Fe_x$ films originate from a single ferromagnetic semiconductor phase, and there is no second phase. This conclusion is the same as that in the previous study on $Ge_{1-x}Fe_x$ films[10].

## III. HOLE TRANSPORT AND BAND PROFILES

### A. Temperature dependence of resistivity and hopping conduction

Figure 3 shows the temperature dependence of the resistivity $\rho$ measured for the undoped $Ge_{1-x}Fe_x$ (blue solid curves) and B-doped $Ge_{1-x}Fe_x$:B (green broken curves) films, in which $\rho$ of Ge:B (black broken curves, sample B0) is also shown in the graph of the samples with $x$ = 2.3 %. In all the undoped $Ge_{1-x}Fe_x$ films, $\rho$ increases with decreasing temperature, and it decreases with increasing $x$ in the $x$ range from 2.3 to 14.0%. On the other hand, $\rho$ of the $Ge_{1-x}Fe_x$:B films with $x$ = 2.3 % is independent of temperature, whereas $\rho$ of the $Ge_{1-x}Fe_x$:B films with $x$ = 6.5, 10.5, and 14.0 % increases with decreasing temperature. Moreover, $\rho$ of the $Ge_{1-x}Fe_x$:B films with $x$ = 6.5, 10.5, and 14.0 % increases with increasing $x$. Comparing the features of both undoped and B-doped films, $\rho$ of the $Ge_{1-x}Fe_x$:B films with higher $x$ (= 6.5, 10.5, and 14.0 %) seems to have the temperature dependence similar to that of the undoped $Ge_{1-x}Fe_x$ films at the same $x$, and $\rho$ of both films are comparable at $x$ = 14.0%. The Ge:B film is degenerate and its $\rho$ is independent of temperature, and its $\rho$ value is smaller than that of the $Ge_{1-x}Fe_x$:B films with $x$ = 2.3 %. These results suggest that the Fermi energy becomes higher and the concentration of conduction holes become smaller with increasing the Fe concentration $x$, and that the effect of B doping is almost negligible at $x$ = 14.0%. We will show the band profiles of $Ge_{1-x}Fe_x$ later in Section III E and F based on our experimental results.

**B. Fitting by the Efros-Shklovskii variable range hopping model**

Since $\rho$ of the undoped Ge$_{1-x}$Fe$_x$ films decreases with increasing temperature and decreases with increasing $x$ in the range from $x = 2.3$ to $14.0$ %, we expect that the transport is dominated by the hopping of itinerant holes between hopping sites which were introduced by the Fe atoms. To confirm this scenario, the conductivity $\sigma$ was plotted as a function of $T^{-1/2}$ and then it was fitted by the following equation based on the Efros-Shklovskii variable range hopping (ES-VRH) model[17,18,19]:

$$\sigma(T) \propto \exp\left[-\left(\frac{T_0}{T}\right)^{\frac{1}{2}}\right], \quad (1)$$

$$T_0 = \frac{\beta e^2}{4\pi\varepsilon_0\varepsilon_r k_B a}, \quad (2)$$

where $T_0$ is a characteristic temperature, $\varepsilon_0\varepsilon_r$ is the dielectric constant of the material, $k_B$ is the Boltzmann constant, $a$ is the localization length, $\beta$ is a dimension-dependent coefficient ($\beta = 2.8$ for three dimensional systems), and $e$ is the elementary charge. Here, it is assumed that $\varepsilon_0\varepsilon_r$ is the same as that of Ge for all the $x$. Figure 4 (a) shows $\sigma - T^{-1/2}$ plots of the undoped Ge$_{1-x}$Fe$_x$ films with $x = 2.3 - 14.0$ %, where open circles and solid lines represent experimental data and fitting, respectively. As can be seen in the figure, the experimental data for all the undoped Ge$_{1-x}$Fe$_x$ films are well fitted by Eq. (1) above $T^{-1/2} = 0.1$ ($T < 100$ K), and thus the transport is dominated by hopping in this temperature range, as expected. Figures 4 (b) and (c) show $T_0$ and $a$ estimated by the fitting, respectively. In Fig. 4 (b), when $x$ increases from $x = 2.3$ to $14.0$ %, $T_0$ monotonically decreases. This indicates that the hopping probability between the sites increases as $x$ increases, owing to the successive change in the material properties. Consequently, $a$ in Fig. 4 (c) increases with increasing $x$. Since the local Fe concentrations fluctuate in the Ge$_{1-x}$Fe$_x$ films and the size of the Fe-rich regions become larger with increasing $x$[13], the increase of $a$ and the decrease of $T_0$ mean that the increase of the average size of the Fe rich regions, which corresponds to $a$, leads to the increase in the hopping probability. Since undoped Ge$_{1-x}$Fe$_x$ films with higher $x$ has higher $\sigma$ as seen in Fig. 3, the transport model is as follows; the Fe-rich regions are the hole hopping sites, and the increase of the average size of the Fe-rich regions leads to the decrease of the average length between the hopping sites, which results in the increase of the hopping probability.

Figure 4(d) shows $\sigma$ vs $T^{-1/2}$ of the Ge$_{1-x}$Fe$_x$:B films, where open circles and solid lines represent experimental data and fitting, respectively. Contrary to the case of the undoped Ge$_{1-x}$Fe$_x$ films, the fitting function of Eq.(1) does not fit to the experimental data of $x = 6.5, 10.5$, and $14.0$ % in the lower $T^{-1/2}$ range; $T^{-1/2} < \sim 0.3$ ($T > \sim 11$ K) for $x = 6.5$ %, $T^{-1/2} < \sim 0.15$ ($T > \sim 45$ K) for $x = 10.5$ %, and $T^{-1/2} < \sim 0.1$ ($T > \sim 100$ K) for $x = 14.0$ %. These fittings by Eq. (1) do not agree well with the experimental data, in contrast to the results of the undoped Ge$_{1-x}$Fe$_x$ films in Fig. 4(a), indicating that another mechanism which is different from variable range hopping contributes to the transport in these lower $T^{-1/2}$ ranges.

Most probably, itinerant holes supplied from the B acceptors contribute to the transport. At $x$ = 2.3 %, $\sigma$ is almost independent of temperature. This strongly indicates that the transport is dominated by the itinerant holes supplied from the B acceptors, namely, $E_F$ is located in the valence band, as is the case of the degenerate p$^+$-type Ge:B film.

**C. Hole concentration $p$ obtained from Hall measurements**

Hall measurements were performed for the undoped $Ge_{1-x}Fe_x$, $Ge_{1-x}Fe_x$:B, and Ge:B films using patterned Hall bar devices (length: 200 μm, width: 50 μm) in the temperature range of 5 – 300 K while a magnetic field was applied perpendicular to the film plane and swept between ±1 T (at $x$ = 2.3 and 6.5 %) or ±5 T (at $x$ = 10.5 and 14.0 %). At $x$ = 2.3 and 6.5 %, the Hall voltage of $Ge_{1-x}Fe_x$ is dominated by the ordinary Hall effect, then the hole concentration $p$ was estimated by the ordinary Hall coefficient $R_H$ at 1 T. In this study, $p$ was defined as $1/eR_H$ for the $Ge_{1-x}Fe_x$ and $Ge_{1-x}Fe_x$:B films. At $x$ = 10.5 and 14.0 %, however, the anomalous Hall effect is not negligible in the Hall voltage, therefore, $p$ was estimated by $R_H$ at 5 T, assuming that the magnetization of a film is saturated and the anomalous Hall effect does not contribute to the change in the Hall voltage (see Section II in SM). Figure 5(a) shows $p$ of the undoped $Ge_{1-x}Fe_x$ films, and $p$ at 300 K in this figure is listed in Table I. As shown in Fig. 5(a), $p$ in the whole temperature range is significantly increased with increasing $x$: $p$ for $x$ = 14.0 % is $3.6 \times 10^{19}$ cm$^{-3}$ at 300 K, which is 30 times larger than that for $x$ = 2.3 %, $1.2 \times 10^{18}$ cm$^{-3}$. As temperature increases, $p$ decreases for $x$ = 10.5 and 14.0 %, whereas it increases for $x$ = 2.3 and 6.5 %. Since the total change in $p$ below 100 K is less than one order of magnitude, it is confirmed that the change in $\sigma$ in Fig. 4(a), which is ~3 – 5 orders of magnitude, is mostly dominated by the hopping probability determined by temperature.

Figure 5(b) shows $p$ of the $Ge_{1-x}Fe_x$:B films, in which the result for the Ge:B film is also plotted as a reference. In contrast to the undoped $Ge_{1-x}Fe_x$ films, $p$ does not significantly vary with $x$ in the $Ge_{1-x}Fe_x$:B films: $p$ decreases with increasing $x$ in the $x$ range of 2.3 – 6.5 %, it shows the lowest value ~$8 \times 10^{18}$ cm$^{-3}$ at 6.5 %, then it increases with increasing $x$ in the $x$ range of 10.5 – 14.0 %, and finally it shows the maximum value ~$1 \times 10^{19}$ cm$^{-3}$ at 14.0 %. As temperature increases, $p$ is almost constant for the Ge:B film, it slightly decreases for $x$ = 2.3 %, and it slightly increases for $x$ = 6.5, 10.5, and 14.0 %. This feature indicates that the dominant hole transport mechanism depends on $x$: At $x$ = 2.3 %, the hole conduction through boron-doped Ge regions (Fe-poor regions) in the $Ge_{1-x}Fe_x$:B film probably dominates the transport, since $\rho$, $\sigma$, and $p$ are almost constant and thus the ES-VRH model cannot be applied in the whole temperature range. At $x$ = 6.5, 10.5, and 14.0 %, both the VRH and the conduction though the Fe-poor regions contribute to the transport in the lower temperature range ($T$ < ~11 K for $x$ = 6.5 %, $T$ < ~45 K for $x$ = 10.5 %, and $T$ < ~100 K for $x$ = 14.0 %), since Eq. (1) is not exactly but roughly fitted to the experimental data in these temperature ranges (Fig. 4(d)). From the results of Figs. 4(d) and 5(b), it is probable that the increase of $x$ reduces the conduction through the Fe-poor regions, which results in the decrease of $p$ in the $x$ range of 2.3 – 6.5 %, but the further increase of $x$ increases the hopping conduction between the Fe-rich regions, which results in the increase of $p$ in the $x$ range of 6.5 – 14.0 %. Thus, the competition of these two mechanisms determines $p$.

**D. Hole concentrations $p$ vs Fe concentration $x$**

To see the relation $p$ versus $x$ in Figs. 5(a) and (b) more in detail, $p$ at 100 K was plotted in Fig. 6(a), where $p$ of the undoped $Ge_{1-x}Fe_x$ and $Ge_{1-x}Fe_x$:B films are represented by the blue circles and green squares, respectively. The difference in $p - x$ relation between the undoped $Ge_{1-x}Fe_x$ and $Ge_{1-x}Fe_x$:B films is clear; the change of $p$ in the undoped $Ge_{1-x}Fe_x$ films is more than two orders of magnitude, whereas that in the $Ge_{1-x}Fe_x$:B films is less than one order of magnitude. Assuming that the holes in the undoped $Ge_{1-x}Fe_x$ films are supplied only by the Fe atoms, the activation rate $p$/Fe atom, namely how many holes are supplied by one Fe atom, was estimated and plotted in Fig. 6(b). Although the activation rate increases with increasing $x$, indicating that the Fe atoms supply holes, it is very small, only 0.04 – 2 %. In the following, we discuss this reason.

Our previous studies using the c-RBS and c-PIXE revealed that about 80 % of the Fe atoms in $Ge_{1-x}Fe_x$ films ($x$ = 6.5 %) are located in the substitutional sites of the diamond-type crystal structure[9], and the XMCD signals indicated that these substitutional Fe atoms in the $Ge_{1-x}Fe_x$ films ($x$ = 6.5 %) are divalent ($Fe^{2+}$)[13]. From these results, a large amount of Fe atoms in the $Ge_{1-x}Fe_x$ films act as acceptors to supply holes, and the values in Fig. 6(b) would not be the activation rate of all the Fe atoms. This probably comes from the fact that the Hall voltage originates from the average hole density in the conduction path, namely, only the hopping holes contributing to the transport induce the Hall voltage. In another previous study, the cross-sectional TEM observation and EDX measurements of $Ge_{1-x}Fe_x$ films ($x$ = 9.5 %) revealed that the local Fe concentration in the Fe-rich regions was $x$ = ~12 % which is significantly higher than that in the Fe-poor regions ($x$ = ~4 %)[8]. From these previous results and the present result showing that the Fe-rich and Fe-poor regions are the hopping sites and the conduction paths, respectively, we conclude that the most holes are localized in the Fe-rich regions. Furthermore, the average hole concentration in the Fe rich regions can be estimated to be more than $10^{20}$ cm$^{-3}$, which is larger than the estimated $p$ in the $Ge_{1-x}Fe_x$ films with $x$ = 14.0 % in Fig. 6(a).

**E. Band profile of the undoped $Ge_{1-x}Fe_x$ films**

The results and discussions described above allow us to illustrate the band profiles of the undoped and B-doped $Ge_{1-x}Fe_x$ films ($x$ = 2.3 – 14 %). Figures 7 (a) and (b) show schematic plan-views of the Fe distribution in the undoped $Ge_{1-x}Fe_x$ films with (a) low Fe content ($x$ = 2.3 %) and (b) high Fe content ($x$ = 6.5, 10.5, 14 %), respectively, where the orange color strength represents the Fe concentration. As $x$ increases, the diameter and density of the Fe-rich regions (the deep orange region) increase, which are obtained from the localization length in Figs. 4 (a) – (c) and our previous results of TEM and EDX[8,9]. In response to these increases, the Fe concentration in the Fe-poor region (the white and pale orange regions) also increases. Figures 7 (c) and (d) show the band profiles of the undoped $Ge_{1-x}Fe_x$ films with $x$ = 2.3 and 10.5 %, respectively, along the broken lines in Figs. 7(a) and (b), respectively. In the figure, $E_F$, C. B. and V. B. denote the Fermi energy, the conduction band bottom, and the valence band top, respectively. The Fe impurity band is located in the band gap at ~0.35 eV above the top of the valence band as revealed by our previous study using ARPES.[15] Since the average hole concentration in the Fe-rich regions is probably more than $10^{20}$ cm$^{-3}$ as described in Section III D, $E_F$ in Figs. 7(c) and (d) is located in the Fe impurity band. In the case of $x$ = 2.3% in Fig. 7(c), since the Fe concentration in the Fe-poor regions is too low to pin $E_F$ in the Fe impurity band and the distance between the Fe-rich regions is longer than the average diameter of the Fe-rich regions, the Fe-poor regions are depleted (there are almost no holes). This results in the relatively low hopping probability for the holes between the hopping sites (the Fe-rich regions), which corresponds to the result of $T_0$ = 1200 K at $x$ = 2.3 % in Fig. 4(b). On the other hand, in the case of $x$ = 10.5% in Fig. 7(b), since the Fe concentration in the Fe-poor regions increases and the distance between the Fe-rich regions decreases compared with the case of $x$ = 2.3%, the Fe poor regions are not depleted and $E_F$ approaches the Fe impurity band. This results in the increase of the hopping probability for the holes, which leads to the reduction of $T_0$ with increasing $x$ as shown in Fig. 4(b).

**F. Band profiles of the $Ge_{1-x}Fe_x$:B films**

We also present the band profiles of the $Ge_{1-x}Fe_x$:B films ($x$ = 2.3 – 14 %) in Fig. 8. Since the properties of the Fe-rich regions, such as the diameter and density, are probably not changed by B doping (this issue will be discussed further in Section IV), the difference in the band profile between the undoped $Ge_{1-x}Fe_x$ and $Ge_{1-x}Fe_x$:B films comes from the properties of the Fe-poor regions. As previously discussed using Figs. 4(d), 5(b), and 6(a), the dominant transport mechanism changes from the conduction through the Fe-poor regions (the Ge:B regions) to the hole hopping between the Fe-rich regions as $x$ increases. Besides, from Figs. 3 and 4(d), it is expected that the band profile of the $Ge_{1-x}Fe_x$:B film with $x$ = 14 % is very similar to that of the undoped $Ge_{1-x}Fe_x$ film with $x$ = 14 %. Figures 8(a) and (b) show schematic plan-views of the Fe distribution in the $Ge_{1-x}Fe_x$:B films with low Fe content ($x$ = 2.3 %) and high Fe content ($x$ = 6.5, 10.5, 14.5 %), respectively. Here the orange and blue color strengths represent the Fe and valence band hole concentrations, respectively. The features of the Fe-rich regions (the deep orange regions) in each figure are the same as those in Figs. 7(a) and (b). At $x$ = 2.3 % (low Fe content), since the conduction through the Fe-poor region, which is nearly heavily-boron-doped Ge:B, is dominant, this region is continuous, which is represented by the continuous blue color as shown in Fig. 8(a). In this continuous region, the pale and white color regions are also drawn, which originates from the Fe atoms, since $p$ in Fig. 5(b) is lower than that of the Ge:B film. At $x$ = 10.5 % (high Fe content), since the hole hopping between the Fe-rich regions is dominant, the degenerated p-type regions are discontinuous which are represented the blue-island regions as shown in Fig. 8(b). Even though they are discontinuous, the conduction through the blue islands in Fig. 8(b) contributes to the transport, but the fraction of this conduction decreases with increasing $x$ because the average size of the blue islands and the distance between the orange islands decrease with increasing $x$.

Figures 8(c) and (d) present the band profiles of the $Ge_{1-x}Fe_x$:B films with $x$ = 2.3 and 10.5 %, respectively, along the respective broken lines in Figs. 8(a) and (b). In the Fe-rich regions, $E_F$ is located in the Fe impurity band in the bandgap, which is the same as that in the undoped $Ge_{1-x}Fe_x$ films (Figs. 7(c) and (d)). At $x$ = 2.3 % (low Fe content), the distance between the Fe-rich regions is relatively large, which leads to the suppression of the hole hopping between the Fe-rich regions. As for the Fe-poor regions, the band profile is significantly different from that of the undoped $Ge_{1-x}Fe_x$ film with $x$ = 2.3 % in Fig. 7 (c): Some regions are degenerated as in the Ge:B film, and other regions are depleted (there are almost no holes in the valence band). The former and latter have relatively lower and higher Fe concentrations, respectively, which comes from the spatial fluctuation of the Fe concentration. Since the degenerated region (blue color) is continuous, a major part of holes transports through this region. On the other hand, when $x$ is increased (6.5, 10.5, and 14.0 %), the degenerated region is divided into some discontinuous regions due to the increase of the depleted regions as $x$ increases, as shown in Fig. 8(b) and (d). In this band profile, the environment surrounding the Fe impurity band holes at $E_F$ in the B-doped $Ge_{1-x}Fe_x$:B film with $x$ = 14.0 % is similar to that of the undoped $Ge_{1-x}Fe_x$ film with $x$ = 14.0 % in Fig. 7(d). Thus, the contribution of the conduction through the Fe-poor regions decreases and that of the hole hopping between the Fe-rich regions in the Fe impurity band increases, and the whole transport properties are similar to those of the undoped $Ge_{1-x}Fe_x$ film ($x$ = 14.0 %). This explains the very similar $\rho - T$ curves of the undoped and B-doped $Ge_{1-x}Fe_x$ films with $x$ = 14.0 % shown in Fig. 3.

Let us briefly summarize the results in Section III. We have shown the temperature dependence of resistivity (Fig. 3) and the relation between the Fe concentration $x$ and hole concentration $p$ (defined as $1/eR_H$) of the $Ge_{1-x}Fe_x$ and $Ge_{1-x}Fe_x$:B films (Fig. 6). Our analysis of the temperature dependence of $\rho$ in undoped $Ge_{1-x}Fe_x$ films (Fig. 4(a) – (c)) indicates that hopping transport in the impurity band of Fe is dominant. To be consistent with our experimental results (Figs. 3 – 6) and previous studies[13,15], we have deduced the band profiles for both of the $Ge_{1-x}Fe_x$ (Figs. 7) and $Ge_{1-x}Fe_x$:B (Figs. 8) films. In the following section, we will discuss the magnetic properties of the $Ge_{1-x}Fe_x$ on the basis of the band profiles (Figs. 7 and 8).

## IV. RELATION BETWEEN THE MAGNETIC PROPERTIES AND HOLE TRANSPORT IN $Ge_{1-x}Fe_x$

In this section, we discuss the magnetic properties of the $Ge_{1-x}Fe_x$ on the basis of the transport properties and band profiles shown in Section III. Figure 9 (a) shows the relation between $T_C$ and $x$ of the undoped $Ge_{1-x}Fe_x$ films (blue circles) and the B-doped $Ge_{1-x}Fe_x$:B films (green squares), where $T_C$ was estimated by the Arrott plots of MCD-$H$ curves as stated in Section II of the main text and Section I of SM. In Fig. 9 (a), $T_C$ increases by 100 K as $x$ is increased from 2.3 to 14.0 % for both the undoped $Ge_{1-x}Fe_x$ and $Ge_{1-x}Fe_x$:B films, and it is seemingly determined by $x$. Thus, $T_C$ is probably related to the increase of the local Fe concentration and the average size of the Fe-rich regions with increasing $x$, but B doping has no influence on $T_C$.

For further analysis, $T_C$ was also plotted as a function of the hole concentration $p$ at 100 K (Fig. 6(a)), as shown in Fig. 9(b), where blue circles and green squares are the results for the undoped $Ge_{1-x}Fe_x$ and $Ge_{1-x}Fe_x$:B films, respectively. When $T_C$ rises by 100 K with increasing $x$ from 2.3 to 14.0 %, $p$ of the undoped $Ge_{1-x}Fe_x$ film monotonically increases from $10^{18}$ to $10^{20}$ cm$^{-3}$, whereas $p$ (=$1/eR_H$) of the $Ge_{1-x}Fe_x$:B film is nearly unchanged at around $1 \times 10^{19}$ cm$^{-3}$. In Section III F, we concluded that $p$ of the $Ge_{1-x}Fe_x$:B films is strongly related to the hole concentration in the Fe-poor regions: $p$ at low $x$ (= 2.3 %) is almost determined by the hole concentration in the Fe-poor regions and $p$ at high $x$ (= 6.5, 10.5 and 14.0 %) is determined by the hole concentrations in the both of Fe-poor and Fe-rich regions. Thus, if the ferromagnetic ordering in both the undoped $Ge_{1-x}Fe_x$ and $Ge_{1-x}Fe_x$:B films would be induced by the holes transporting through the Fe-poor regions, $T_C$ of the $Ge_{1-x}Fe_x$:B film would become higher than that of the undoped $Ge_{1-x}Fe_x$ film with the same $x$ (= 2.3 – 14.0 %). From this consideration, Fig. 9(b) means that the ferromagnetic ordering in the $Ge_{1-x}Fe_x$ and $Ge_{1-x}Fe_x$:B films is not induced by the holes in the Fe-poor regions. On the other hand, since the average distance between the Fe atoms is 2 – 4 times larger than the nearest-neighbor distance between the Ge atoms in the given Fe content $x$, the direct ferromagnetic exchange interaction between the Fe atoms is hardly induced. Thus, it is quite reasonable to conclude that the holes in the Fe-rich regions are strongly related to the ferromagnetic ordering in both the $Ge_{1-x}Fe_x$ and $Ge_{1-x}Fe_x$:B films. Based on this consideration on the data of Fig. 9 together with all the results described in the previous sections, we discuss the origin of the ferromagnetic ordering of the $Ge_{1-x}Fe_x$ and $Ge_{1-x}Fe_x$:B films in the following.

To discuss the origin of the ferromagnetic ordering, the band profiles in Figs. 7 and 8 are very helpful. In both films, $E_F$ is located in the Fe impurity band which is formed by the Fe-rich regions and the average spatial size of the Fe impurity band becomes larger with increasing $x$ as shown in Figs. 7 and 8. Since the hole density of the Fe impurity band is probably more than $10^{20}$ cm$^{-3}$ as discussed in Section III D, such a large number of holes hopping in the Fe impurity band can induce the ferromagnetic ordering between the Fe atoms via the double exchange interaction.[15] The strength of the ferromagnetic ordering, which is reflected in $T_C$, is determined by the total number of spins in one Fe impurity band, which becomes larger with increasing $x$. In the case of the $Ge_{1-x}Fe_x$:B films at high $x$ (6.5, 10.5, and 14 %), since $E_F$ also intersects the valence band in the Fe-poor regions (Fig. 8 (d)), a large number of holes can also transport from one Fe impurity band to another through the valence band. However, the holes may lose their spin information during the transport, and thus ferromagnetic ordering between the Fe-rich regions is not induced at relatively high temperature. Thus, the region of the ferromagnetic ordering almost overlaps with the Fe-rich region and $T_C$ is determined only by $x$. As shown in Fig. 9(a), B doping has no influence on $T_C$. This is because the environment surrounding the impurity band holes in the Fe-rich region are not changed by B doping (Fig 7(d) and 8(d)) as discussed above. Our discussion in this section is consistent with the conclusion in the previous studies[9,13]: The ferromagnetic ordering occurs around the Fe-rich regions and it is more pronounced by the increase in the fluctuation of the spatial Fe concentration between the Fe-poor and Fe-rich regions.

## V. OTHER TRANSPORT PROPERTIES OF $Ge_{1-x}Fe_x$

### A. Anomalous Hall conductivity *vs* conductivity

In this section, we show additional experimental results of transport properties in the undoped $Ge_{1-x}Fe_x$ ($y = 0$) and $Ge_{1-x}Fe_x$:B ($y = 4.4 \times 10^{19}$ cm$^{-3}$) films. To investigate the origin of the anomalous Hall effect of the undoped $Ge_{1-x}Fe_x$ and $Ge_{1-x}Fe_x$:B films with $x$ = 10.5 and 14.0 %, we performed Hall measurements using a cryostat equipped with a superconducting magnet under a magnetic field up to 5 T. The anomalous Hall effect with clear hysteresis loops was observed (see Section II of SM). In the anomalous Hall effect data, the influence of magnetoresistance is negligible, since the magnetoresistance of the undoped $Ge_{1-x}Fe_x$ and $Ge_{1-x}Fe_x$:B films is negligibly small (less than 1.0 % at 5 T). Figure 10 shows the anomalous Hall conductivity $\sigma_{AHE}$ as a function of the conductivity $\sigma$ of the $Ge_{1-x}Fe_x$ and $Ge_{1-x}Fe_x$:B films. Here, $\sigma_{AHE}$ was estimated from the Hall resistance data (see Section II of SM). The value of scaling parameter $\gamma$ in the relation $\sigma_{AHE} \propto \sigma^{\gamma}$ was obtained by fitting in Fig. 10: $\gamma$ of the $Ge_{1-x}Fe_x$ films with $x$ = 10.5 and 14.0 % was estimated to be 1.0 and 1.2, and $\gamma$ of the $Ge_{1-x}Fe_x$:B films with $x$ = 10.5 and 14.0 % was estimated to be 1.5 and 1.4, respectively. The data near $T_C$ (upper right side of the plots in the Fig. 10) were not fitted by the relation $\sigma_{AHE} \propto \sigma^{\gamma}$, since the estimation of $\sigma_{AHE}$ is less accurate in that temperature range when we subtract the ordinary Hall effect contribution from the raw Hall data (section II and III in SM). We analyzed the data in Fig. 10 from the viewpoint of hopping conduction as follows.

The $\gamma$ values of the $Ge_{1-x}Fe_x$ and $Ge_{1-x}Fe_x$:B films with $x$ = 10.5 and 14.0 % are in the range from 1.0 to 1.5, which are similar to the experimentally reported $\gamma$ values around 1.5 for FMSs (e.g. (Ga,Mn)As[20,21]). A theory of the anomalous Hall effect predicts that $\gamma$ is around 1.5 in the hopping transport regime and $\gamma$ values from 1.33 to 1.76 in the ES-VRH regime[22] in insulating materials such as FMSs. Since the $\gamma$ values estimated by the data in Fig. 10 are in the range of 1.0 − 1.5, this theory can also be applied to our $Ge_{1-x}Fe_x$ and $Ge_{1-x}Fe_x$:B films. The carrier transport in the $Ge_{1-x}Fe_x$ and $Ge_{1-x}Fe_x$:B films with $x$ = 10.5 and 14.0 % are explained by the ES-VRH regime as described in Section III

B. From this point of view, the hole hopping between the Fe-rich regions in the Fe impurity band dominantly contribute to the anomalous Hall effect: In the $Ge_{1-x}Fe_x$ and $Ge_{1-x}Fe_x$:B films at $x$ = 10.5 and 14.0 %, the holes transport by the hopping through the impurity band of Fe formed by the Fe-rich regions (Figs. 7(b)(d) and 8(b)(d)), as discussed in Section III E and F and Section IV.

## B. Temperature dependence of the hole mobility $\mu$

The temperature dependences of the hole mobility $\mu$ are shown in figures 11(a) and (b) for the undoped $Ge_{1-x}Fe_x$ and $Ge_{1-x}Fe_x$:B films with $x$ = 2.3 – 14.0 %, respectively, where each $\mu = 1/ep\rho$ was obtained from the temperature dependences of $\rho$ in Fig. 3 and $p$ in Fig. 5. As shown in Figs. 11(a) and (b), $\mu$ in both the $Ge_{1-x}Fe_x$ and $Ge_{1-x}Fe_x$:B films decrease with increasing $x$, which is similar to the B doping concentration dependence of $\mu$ in Ge with a high B doping range (~$10^{19}$ – $10^{21}$ cm$^{-3}$).[23] We also found that $\mu$ of each $Ge_{1-x}Fe_x$ film in Fig. 11(a) is lower than $\mu$ of each $Ge_{1-x}Fe_x$:B film with the same $x$ in Fig. 11(b) at the same temperature, and that in all the $Ge_{1-x}Fe_x$ films, $\mu$ increases with increasing temperature. This feature can be explained by the hole transport discussed in section Section III E and F: The transport in the undoped $Ge_{1-x}Fe_x$ films is dominated by the hopping through the Fe impurity band. In contrast, the transport in the $Ge_{1-x}Fe_x$:B films comes from both the Fe impurity band and valence band conduction. The $\mu - T$ data for the undoped $Ge_{1-x}Fe_x$ films with $x$ = 2.3 – 14.0 % are roughly fitted by the relation $\mu \propto T^{3/2}$ which is shown by the dotted line in Fig. 11(a), indicating that the dominant scattering mechanism is ionized impurity ($Fe^{2+}$)[13] scattering in the Fe rich regions. We also noticed that $\mu$ of the undoped $Ge_{1-x}Fe_x$ films with $x$ = 2.3 – 14.0 % in Fig. 11(a) is around 0.01 – 10 cm$^2$/sV, which is smaller than $\mu$ (> 100 cm$^2$/sV) of the degenerate semiconductor Ge:B film in Fig. 11(b). The lower $\mu$ supports the hole transport and band profile model of the undoped $Ge_{1-x}Fe_x$ films shown in Fig. 7 (Section III E): In the $Ge_{1-x}Fe_x$ films ($x$ = 2.3 – 14.0 %), $E_F$ is located at the impurity band of Fe and the holes hop via the Fe-rich regions.

On the other hand, $\mu$ of the Ge:B and $Ge_{1-x}Fe_x$:B films with $x$ = 2.3 % is nearly constant and independent of temperature, as shown in Fig. 11(b). This result suggests that the dominant scattering mechanism is neutral impurity scattering in the Fe poor regions. These neutral impurities would be self interstitial Ge atoms or vacancies due to the nonequilibrium MBE growth at 200°C, or dislocations due to the lattice mismatch (~4 %) between the $Ge_{1-x}Fe_x$:B films and SOI(001) substrates. In contrast, $\mu$ of the $Ge_{1-x}Fe_x$:B films with $x$ = 6.5 – 14.0 % slightly increases with increasing temperature. Considering the $\mu-T$ relation of the $Ge_{1-x}Fe_x$ films in Fig. 11(a), the $\mu-T$ relation of the $Ge_{1-x}Fe_x$:B films is probably also caused by ionized impurity ($Fe^{2+}$) scattering in the Fe rich regions. Namely, $\mu$ of the $Ge_{1-x}Fe_x$:B films is determined both by the neutral impurity scattering in the Fe poor regions and by the ionized impurity ($Fe^{2+}$) scattering in the Fe rich regions. This is consistent with the hole transport and band profile model of the $Ge_{1-x}Fe_x$ films shown in Fig. 8 (Section III F): In the $Ge_{1-x}Fe_x$:B films ($x$ = 6.5 – 14.0 %), $E_F$ is located in the Fe impurity band or in the valence band depending on the local Fe concentration, and the holes transport via both the Fe-rich regions and the valence band regions.

In this section, we have shown additional experimental results of transport properties (anomalous Hall conductivity and mobility $\mu$) of the undoped $Ge_{1-x}Fe_x$ and $Ge_{1-x}Fe_x$:B films. Table II summarizes the hole transport properties in the undoped $Ge_{1-x}Fe_x$ and $Ge_{1-x}Fe_x$:B (B concentration $y = 4.4 \times 10^{19}$ cm$^{-3}$) films obtained in this work; whether the temperature dependence of the resistivity is metallic or insulating, whether hopping transport is dominant or not, and the relation between $p$ vs $x$, and the relation between $\mu$ vs $T$. The whole experimental results presented in this paper are explained by the band profiles shown in Figs. 7 and 8 in Section III.

## VI. CONCLUSION

We have clarified the systematic and detailed hole transport and magnetic properties of the $Ge_{1-x}Fe_x$ ($x$ = 2.3 – 14.0 %) films without and with boron (B) doping grown on SOI (001) substrates. Firstly, we have shown the temperature dependence of resistivity and hole concentration in Section III. From the results and analysis, we have illustrated the band profiles of the $Ge_{1-x}Fe_x$ and $Ge_{1-x}Fe_x$:B films: In case of the undoped $Ge_{1-x}Fe_x$ films, the hole transport dominated by hopping in the Fe impurity band between the Fe rich regions which are formed by the nanoscale spatial Fe concentration fluctuations. In the boron-doped $Ge_{1-x}Fe_x$:B films, on the other hand, at low $x$ (= 2.3 %) the hole transport is dominated by the valence band conduction, and at high $x$ (= 6.5, 10.5, 14.0 %) the hole transport is a mixture of the valence band conduction and hopping conduction. Secondly, we showed in Section IV that the magnetic properties of the $Ge_{1-x}Fe_x$ films are not influenced by B-doping at $x$ = 2.3 – 14.0 %. We figured out why the ferromagnetism in the $Ge_{1-x}Fe_x$ films without and with B-doping are almost the same from the band profiles: Because of the high density of states in the Fe-rich regions, the local hole concentration is not affected by B doping. Besides, the holes in the Fe rich regions mediate the ferromagnetic order via the double-exchange mechanism, but the holes in the Fe poor regions do not contribute to the ferromagnetic order. This mechanism can explain all of our experimental results presented in this study, and is consistent with the conclusion in the previous studies, particularly XMCD[13] and SX-ARPES[15]. Thirdly, we showed the additional transport properties (anomalous Hall conductivity and mobility) in Section V. In conclusion, all the results obtained from the transport and the magnetic characteristics indicate that the ferromagnetism in the $Ge_{1-x}Fe_x$ films is closely linked to the nanoscale spatial Fe concentration fluctuations and the hole hopping conduction. Our pictures of the band profiles of the $Ge_{1-x}Fe_x$ films present comprehensive understanding of the hole transport and the origin of ferromagnetism, and will provide important guidance to develop spin transport devices utilizing $Ge_{1-x}Fe_x$.

**ACKNOWLEDGEMENTS**

This work was partly supported by Grants-in-Aid for Scientific Research (22224005, 23000010, 26249039, and 15H02109), including the Specially Promoted Research, CREST of JST, and the Spintronics Research Network of Japan.

**TABLES AND FIGURES**

| Sapmle | Fe content $x$ (%) | Boron conc. $y$ (cm$^{-3}$) | Hole conc. $p$ (cm$^{-3}$) at 300K | Curie temperature $T_C$ (K) |
|---|---|---|---|---|
| A1 | 2.3 | Undoped | $1.2\times10^{18}$ | 15 |
| A2 | 6.5 | Undoped | $2.7\times10^{18}$ | 35 |
| A3 | 10.5 | Undoped | $1.1\times10^{19}$ | 95 |
| A4 | 14.0 | Undoped | $3.6\times10^{19}$ | 110 |
| B0 | 0 | $4.4\times10^{19}$ | $2.8\times10^{19}$ | - |
| B1 | 2.3 | $4.4\times10^{19}$ | $1.7\times10^{19}$ | 17 |
| B2 | 6.5 | $4.4\times10^{19}$ | $1.3\times10^{19}$ | 45 |
| B3 | 10.5 | $4.4\times10^{19}$ | $1.8\times10^{19}$ | 85 |
| B4 | 14.0 | $4.4\times10^{19}$ | $2.4\times10^{19}$ | 115 |

TABLE I. Samples examined in this work and their parameters. Here, $x$ is estimated by Rutherford backscattering spectrometry, $y$ is estimated by secondary ion mass spectrometry, $p$ (=$1/eR_H$) is estimated from Hall measurements, and Curie temperature $T_C$ is estimated from magnetic circular dichroism (MCD) measurements and Arrott plots of the MCD-$H$ curves (see Section I in supplementary material), respectively.

|  | $Ge_{1-x}Fe_x$ ($y = 0$) | $Ge_{1-x}Fe_x$:B ($y = 4.4\times10^{19}$/cm$^3$) |
| --- | --- | --- |
| Resistivity $\rho$ | insulating | $x = 2.3$ % : metallic <br> $x = 6.5 – 14.0$ % : insulating |
| Hole conduction (ES-VRH) | hopping | not dominated by hopping |
| Change in $p$ with increasing $x$ | increase | almost constant |
| Mobility $\mu$ with increasing $T$ | increase | almost constant |

TABLE II. Summary of the hole transport properties in the undoped $Ge_{1-x}Fe_x$ and $Ge_{1-x}Fe_x$:B (B concentration $y = 4.4\times10^{19}$ cm$^{-3}$) films. The table shows whether the temperature dependence of the resistivity is metallic or insulating, whether hopping transport is dominant or not, the relation between $p$ ($=1/eR_H$) vs $x$, and the relation between $\mu$ vs $T$ in $Ge_{1-x}Fe_x$ ($y = 0$, without B) and $Ge_{1-x}Fe_x$:B ($y = 4.4 \times 10^{19}$ cm$^{-3}$).

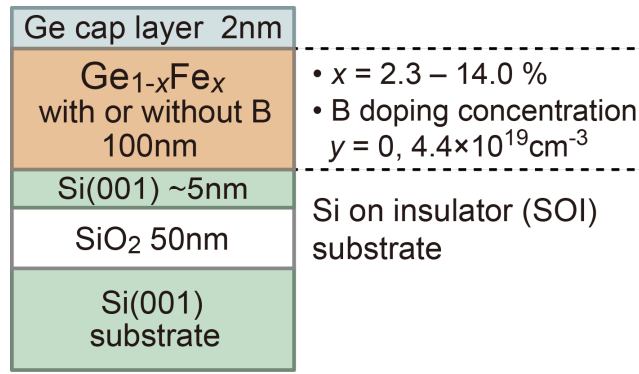

FIG. 1. Schematic structure of the Ge$_{1-x}$Fe$_x$ single layer samples. Ge cap (2nm) and the Ge$_{1-x}$Fe$_x$ (100 nm) layers are grown by MBE on silicon-on-insulator (SOI) substrates with a very thin (~5 nm) Si top layer.

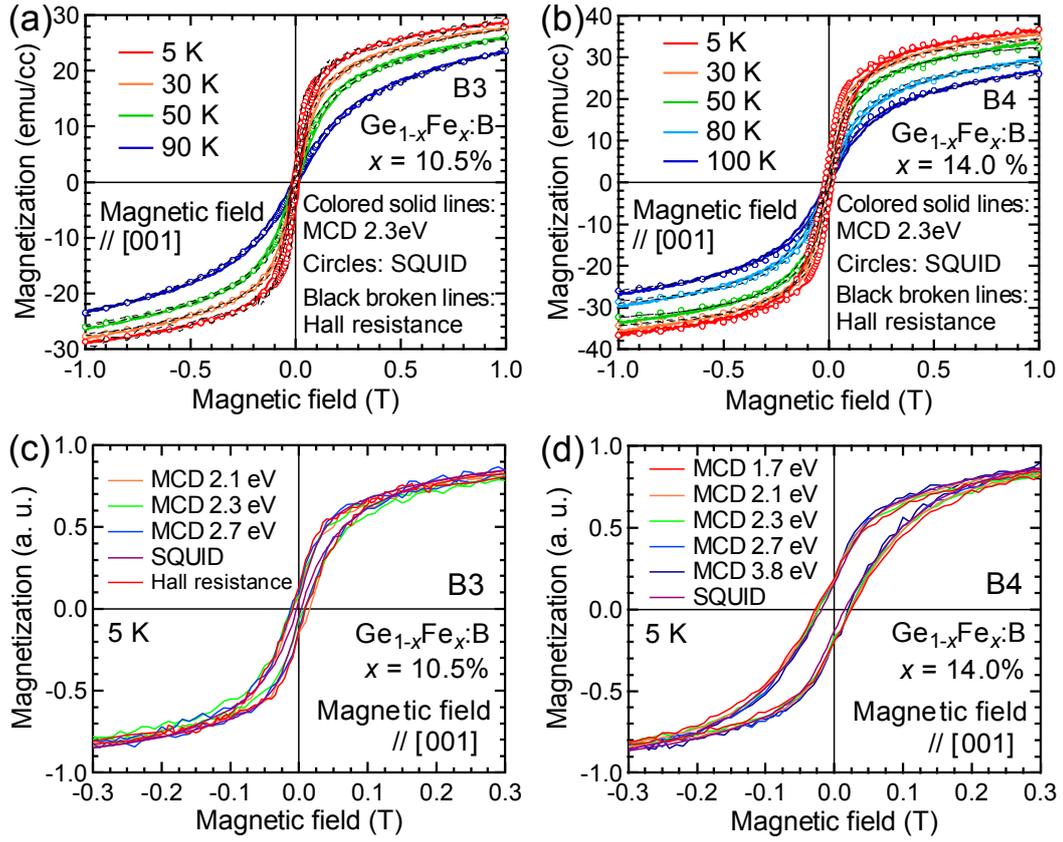

FIG. 2. (a)(b) *M-H* curves of the $Ge_{1-x}Fe_x$:B ($y = 4.4\times10^{19}$ cm$^{-3}$) films at various temperatures, (a) *x* = 10.5 %, (b) *x* = 14.0 %, measured by MCD spectrometry, a Quantum Design MPMS-5S SQUID Magnetometer, and Hall measurements (anomalous Hall effect). (c)(d) Normalized *M-H* curves of the $Ge_{1-x}Fe_x$:B ($y = 4.4\times10^{19}$ cm$^{-3}$) films at 5 K, (c) *x* = 10.5 %, (d) *x* = 14.0 %, measured by MCD spectrometry and the SQUID Magnetometer.

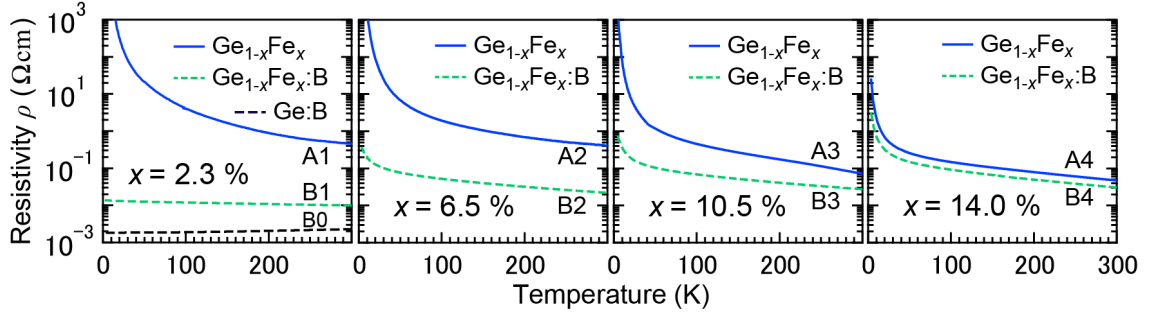

FIG. 3. Temperature dependence of the resistivity $\rho$ of undoped $Ge_{1-x}Fe_x$ ($y = 0$, without B, solid blue curves), and $Ge_{1-x}Fe_x$:B ($y = 4.4 \times 10^{19}$ cm$^{-3}$, broken green curves), and Ge:B ($x = 0$, $y = 4.4 \times 10^{19}$ cm$^{-3}$, broken black curves) films as a reference sample.

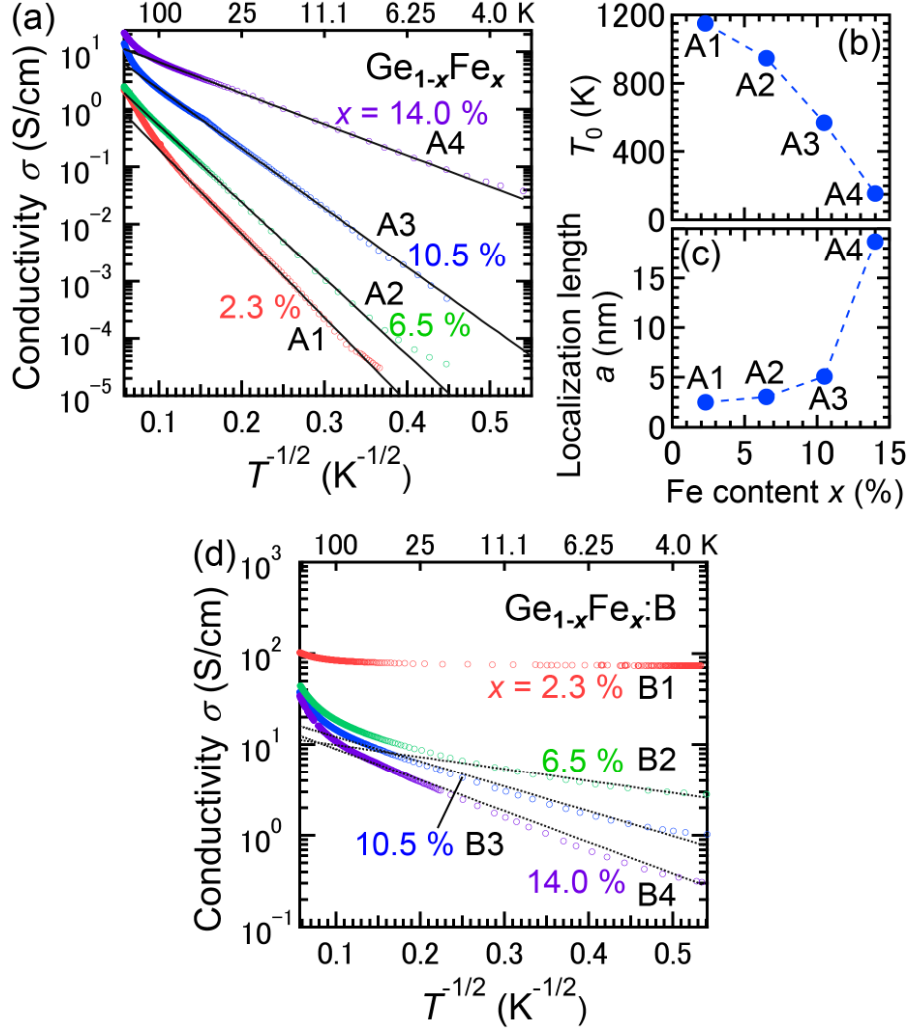

FIG. 4 (a) Conductivity $\sigma$ vs $T^{-1/2}$ plots for undoped $Ge_{1-x}Fe_x$ ($y = 0$) films in the range from 3.5 K to 300 K. Black lines in (a) show ES-VRH model fits. From the fits, $T_0$ and localization length $a$ (in Eq. (1)) are obtained and plotted as a function of $x$, as shown in (b) and (c). (d) $\sigma$ vs $T^{-1/2}$ plots for the $Ge_{1-x}Fe_x$:B ($y = 4.4 \times 10^{19}$ cm$^{-3}$) films in the range from 3.5 K to 300 K. These plots do not agree well with the ES-VRH model fits (black dotted line of (d)) in the high temperature region (low $T^{-1/2}$ region).

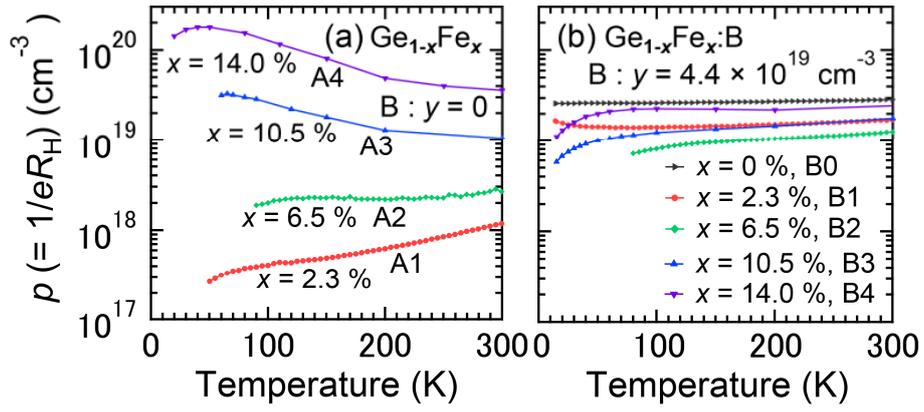

FIG. 5. Temperature dependence of the hole concentration $p$ (=$1/eR_H$) in the (a) $Ge_{1-x}Fe_x$ ($y = 0$) and (b) $Ge_{1-x}Fe_x$:B ($y = 4.4\times10^{19}$ cm$^{-3}$) films at $x$ = 2.3, 6.5, 10.5, and 14.0 %. Here, $p$ was estimated by the ordinary Hall coefficient $R_H$ at 1 T ($x$ = 2.3 and 6.5 %) and at 5 T ($x$ = 10.5, and 14.0 %) (see Section II in supplementary material).

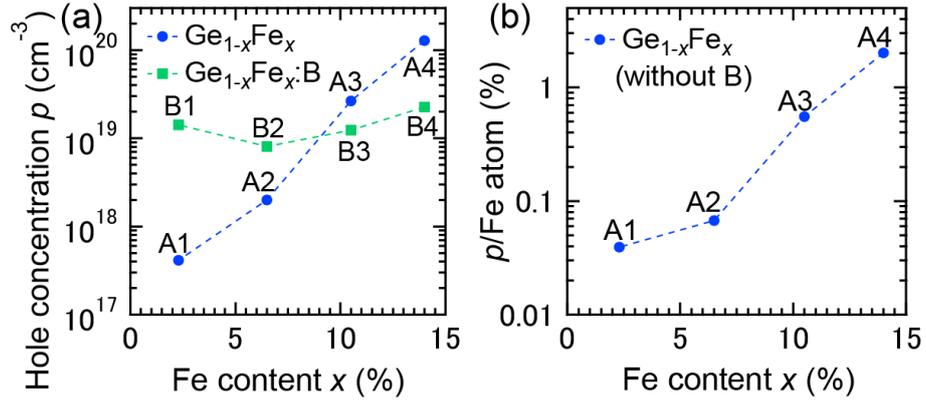

FIG. 6. (a) Hole concentration $p$ (=$1/eR_H$) at 100 K as a function of $x$ of undoped $Ge_{1-x}Fe_x$ ($y = 0$, blue circles) and $Ge_{1-x}Fe_x$:B ($y = 4.4 \times 10^{19}$ cm$^{-3}$, green squares) films. (b) $p$/Fe atom as a function of $x$ of undoped $Ge_{1-x}Fe_x$ ($y = 0$, blue circles) at 100 K.

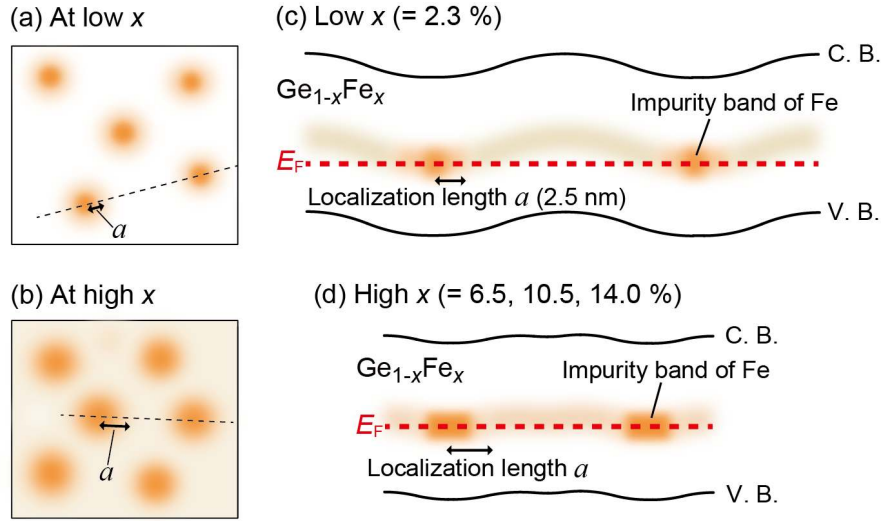

FIG. 7. Schematic real space pictures of local Fe concentration of undoped $Ge_{1-x}Fe_x$ ($y = 0$) films at (a) low and (b) high Fe content $x$, and illustrations of the band profile in the $Ge_{1-x}Fe_x$ ($y = 0$) films at (c) low Fe content ($x = 2.3$ %) and (d) high Fe content ($x = 6.5, 10.5,$ and $14.0$ %). (c) and (d) are depicted along the black dotted lines in (a) and (b). $E_F$, C. B. and V. B. in (c) and (d) denote the Fermi energy, the conduction band bottom, and the valence band top, respectively. Deep orange and pale orange colors in the figures represents the Fe impurity band in Fe-rich and Fe-poor regions, respectively, and the orange color strength represents the Fe concentration. The white color represents the depletion (no carrier) region. The deep orange regions in (a) and (b) are the hopping sites.

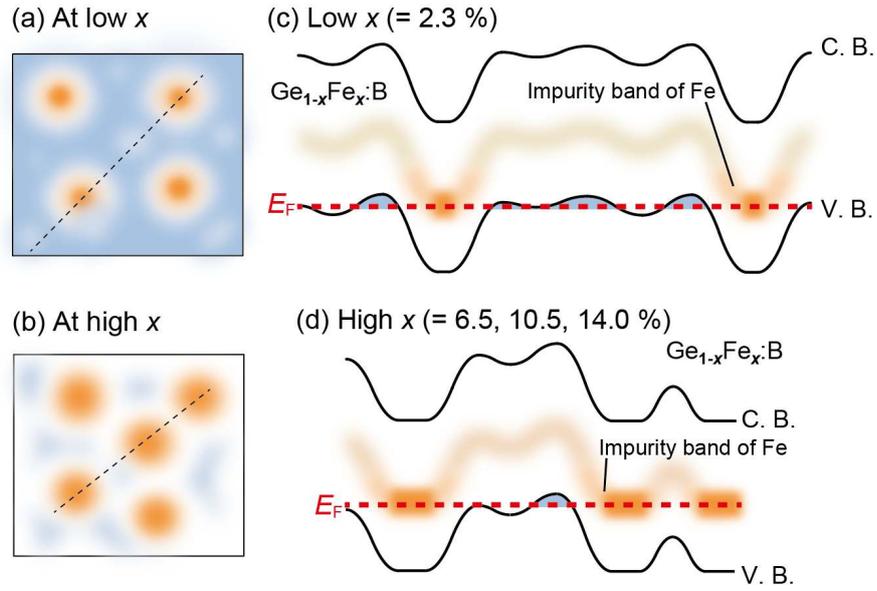

FIG. 8. Schematic real space pictures of the local hole and Fe concentrations at (a) low and (b) high Fe content $x$, and illustrations of the band profiles in boron-doped $Ge_{1-x}Fe_x$:B ($y = 4.4 \times 10^{19}$ cm$^{-3}$) films at (c) low Fe content $x$ (= 2.3 %) and (d) high $x$ (= 6.5, 10.5, and 14.0 %). (c) and (d) are depicted along the black dotted lines in (a) and (b). $E_F$, C. B. and V. B. in (c) and (d) denote the Fermi energy, conduction band bottom, and valence band top, respectively. Deep and pale orange colors in the figures represent the impurity band of Fe in Fe-rich and poor regions, respectively. The orange color strength represents the Fe concentration. Blue color indicates the regions in which $E_F$ lies in the valence band thus there are holes in the valence band, and white color represents the depletion (no carrier) region.

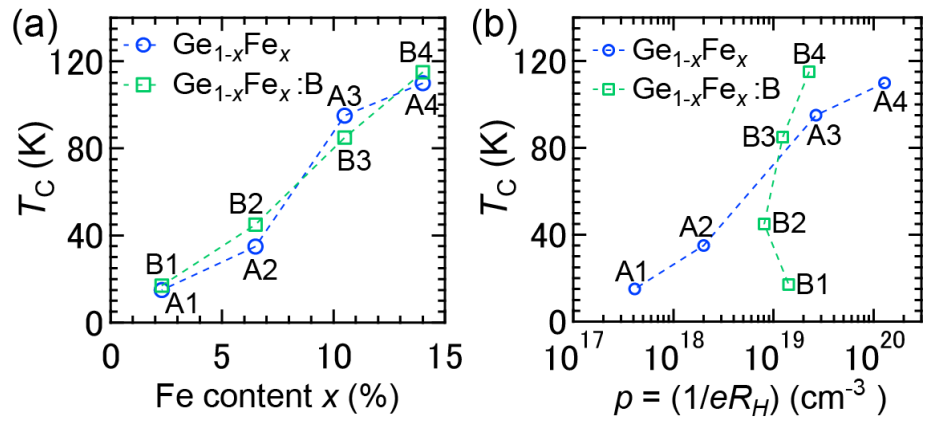

FIG. 9. (a) Relations between the Curie temperature $T_C$ and the Fe content $x$ in undoped $Ge_{1-x}Fe_x$ ($y = 0$, blue circles) and $Ge_{1-x}Fe_x$:B ($y = 4.4 \times 10^{19}$ cm$^{-3}$, green squares) films. (b) Relations of the $T_C$ and the hole concentration $p$ (at 100 K) of undoped $Ge_{1-x}Fe_x$ ($y = 0$, blue circles) and $Ge_{1-x}Fe_x$:B ($y = 4.4 \times 10^{19}$ cm$^{-3}$, green squares) films.

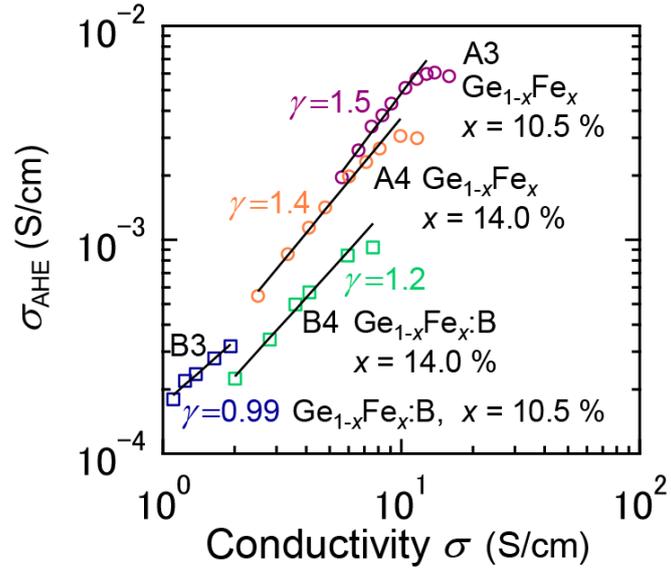

FIG. 10. Anomalous Hall conductivity $\sigma_{AHE}$ vs conductivity $\sigma$ of undoped $Ge_{1-x}Fe_x$ ($y = 0$) and $Ge_{1-x}Fe_x$:B ($y = 4.4 \times 10^{19}$ cm$^{-3}$) films at $x = 10.5, 14.0$ %. $\sigma_{AHE}$ was extracted from the raw data of the Hall resistance measured under a magnetic field from -5 to 5 T at $T = 10 \sim 90$ K (see Section II and III in supplementary material).

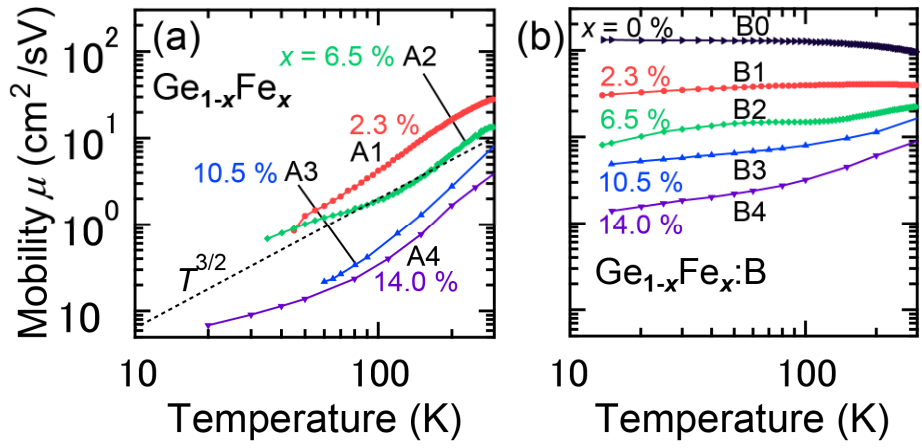

FIG. 11. Temperature dependence of the hole mobility $\mu$ in (a) undoped $Ge_{1-x}Fe_x$ ($y = 0$) and (b) $Ge_{1-x}Fe_x$:B ($y = 4.4\times10^{19}$ $cm^{-3}$) films.

*Supplementary Material*

# Impurity band conduction in group-IV ferromagnetic semiconductor $Ge_{1-x}Fe_x$ with nanoscale fluctuations in Fe concentration

Yoshisuke Ban,[1] Yuki K. Wakabayashi,[1] Ryosho Nakane,[1] and Masaaki Tanaka[1,2]

[1]*Department of Electrical Engineering and Information Systems, The University of Tokyo, 7-3-1 Hongo, Bunkyo-ku, Tokyo 113-8656, Japan*

[2]*Center for Spintronics Research Network, Graduate School of Engineering, The University of Tokyo, 7-3-1 Hongo, Bunkyo-ku, Tokyo 113-8656, Japan*

**Section I. Estimation of the Curie temperature of the undoped $Ge_{1-x}Fe_x$ and boron-doped $Ge_{1-x}Fe_x$:B films by the Arrott plots of the magnetic field dependence of the MCD intensity**

The Curie temperature $T_C$ of the undoped $Ge_{1-x}Fe_x$ and boron-doped $Ge_{1-x}Fe_x$:B ($x$ = 2.3 – 14.0 %) films in Table I was estimated by the Arrott plots of the magnetic field dependence of the reflection magnetic circular dichroism (MCD) intensity (MCD − $H$ curves). The MCD intensity is proportional to the magnetization of the films as presented in Fig. 2, and does not contain the diamagnetic contribution of the SOI substrates. Our previous work of XMCD study[13] showed that the entire film is ferromagnetic at $T < T_C$ and the film becomes superparamagnetic above $T_C$. Fig. S1(a) and (b) show the MCD − $H$ curves of the $Ge_{1-x}Fe_x$ and $Ge_{1-x}Fe_x$:B ($x$ = 14.0 %) films measured at a photon energy of 2.3 eV (the $E_1$ transition energy of bulk Ge) at various temperatures. To estimate $T_C$, we made Arrott plots (MCD)$^2$ − $\mu_0 H$/MCD, as shown in Fig. S1(c) and (d), where MCD is the reflection MCD intensity and $\mu_0 H$ (= $B$) is the magnetic field applied perpendicular to the film plane. From Fig. S1(c) and (d), the Curie temperature $T_C$ of the (c) $Ge_{1-x}Fe_x$ ($x$ = 14.0 %) and (d) $Ge_{1-x}Fe_x$:B ($x$ = 14.0 %) films was estimated to be 110 K and 115 K, respectively.

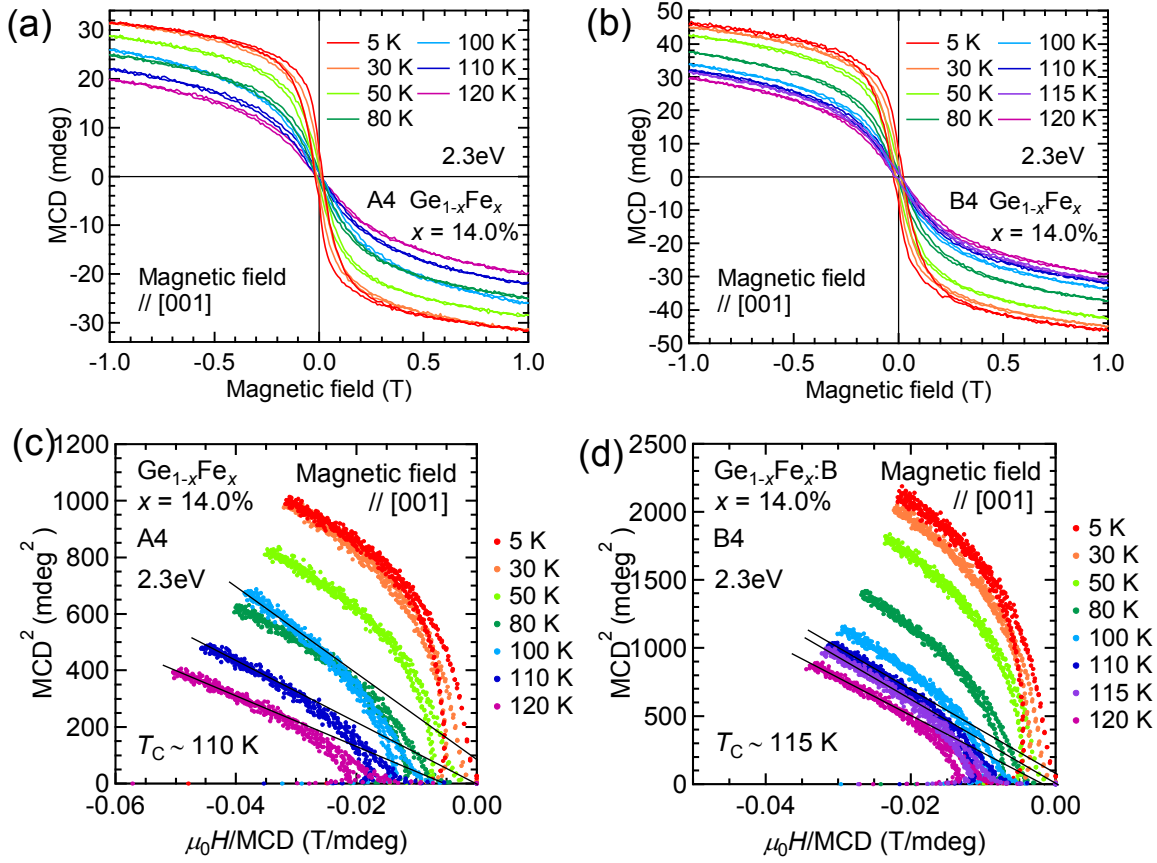

FIG. S1. (a)(b) Magnetic field dependence of the reflection MCD intensity (MCD-$H$ curves) of (a) $Ge_{1-x}Fe_x$ and (b) $Ge_{1-x}Fe_x$:B films at $x$ = 14.0 %. MCD intensity was measured at the temperatures from 5 to 120 K with a perpendicular magnetic field $\mu_0 H$ = -1 ~ 1 T. (c)(d) Arrott plots, $(MCD)^2 - \mu_0 H/MCD$, for the MCD-$H$ curves of the (c) $Ge_{1-x}Fe_x$ ($x$ = 14.0 %) and (d) $Ge_{1-x}Fe_x$:B ($x$ = 14.0 %) films. The MCD intensity was measured at a photon energy of 2.3 eV (the $E_1$ transition energy of bulk Ge).

## Section II. Extraction of the anomalous Hall component from the raw Hall data

The Hall measurements were carried out on the $Ge_{1-x}Fe_x$ ($x$ = 10.5 and 14.0 %, $y$ = 0 and 4.4×10$^{19}$ cm$^{-3}$) films under a magnetic field up to 5 T with a cryostat equipped with a superconducting magnet. We observed the anomalous Hall effect (AHE) with clear hysteresis loops of the $Ge_{1-x}Fe_x$ films, as shown in Fig. S2. The Hall resistance $R_{xy}$ (Fig. S2) is generally given by $R_{xy} = R_H B/d + \rho_{AHE}/d$, where $R_H$, $B$, $d$, and $\rho_{AHE}$ are the ordinary Hall coefficient, the magnetic field, the film thickness, and the anomalous Hall resistivity, respectively. The method to extract the anomalous Hall resistivity $\rho_{AHE}$ from the raw data in Fig. S2 is as follows: Assuming that the magnetization (which is proportional to the anomalous Hall component) of the $Ge_{1-x}Fe_x$ films is saturated at less than 4 T, we subtract the linear slope between 4 and 5 T as the ordinary Hall component from the Hall resistance. Thus the anomalous Hall component $\rho_{AHE}$ is extracted. In this method, we subtract not only the ordinary Hall effect, but also the paramagnetic component as a linear slope. These two components are impossible to separate. From the anomalous Hall resistivity $\rho_{AHE}$ obtained above and resistivity $\rho$, the anomalous Hall conductivity $\sigma_{AHE}$ in Fig. 10 is obtained by $\sigma_{AHE} = \rho_{AHE}/(\rho^2 + \rho_{AHE}^2) \approx \rho_{AHE}/\rho^2$. The underestimation of $\rho_{AHE}$ and $\sigma_{AHE}$ appears especially when temperature becomes close to $T_C$ because the paramagnetic component becomes larger (see Section III in SM). Therefore, the $\gamma$ values in the main text are inevitably underestimated.

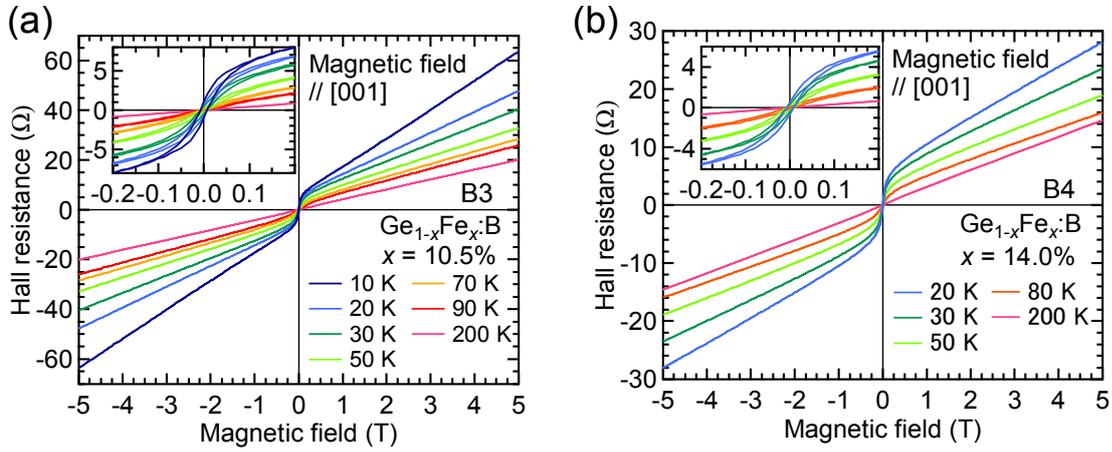

FIG. S2. Magnetic field dependence of the Hall resistance of the $Ge_{1-x}Fe_x$:B ($y$ = 4.4×10$^{19}$ cm$^{-3}$) films at (a) $x$ = 10.5 %, (b) $x$ = 14.0 % at various temperatures.

**Section III. Ratio of the paramagnetic contribution to the ferromagnetic contribution in the total magnetization**

The $\sigma_{AHE}$ vs $\sigma$ plots in the Fig. 10 near $T_C$ (upper right side of the plots) are not fitted by $\sigma_{AHE} \propto \sigma_{xx}^{\gamma}$. This is mainly because of the method of subtracting the ordinary Hall effect, as described in Section II. In this method, linear slopes of the Hall resistance $R_{xy}$ between 4 and 5 T are presumed to be only the ordinary Hall component, but simultaneously, the paramagnetic AHE component is also present and subtracted. Hence, the paramagnetic component cannot be separated from the ordinary Hall effect in this method. The ratio of the paramagnetic contribution to the ferromagnetic contribution in the total magnetization at various temperatures can be estimated from the magnetic field dependence of the Fe selective XMCD intensity shown in Fig. S3 (a)[13]. In the case of the $Ge_{1-x}Fe_x$ ($x$ = 6.5 %) film grown on a Ge (001) substrate[13], the ratio near $T_C$ (= 100 K) becomes larger than that at lower temperature as shown in Fig. S3 (b). This can explain the results of Fig. 10. The difference in $\sigma_{AHE}$ ($\approx \rho_{AHE}/\rho^2$) between the fit and the data in Fig. 10 (upper right side of the plots) is enlarged near $T_C$, because the ratio of the paramagnetic contribution to the linear slope is larger than that at lower temperatures (lower left side of the plots).

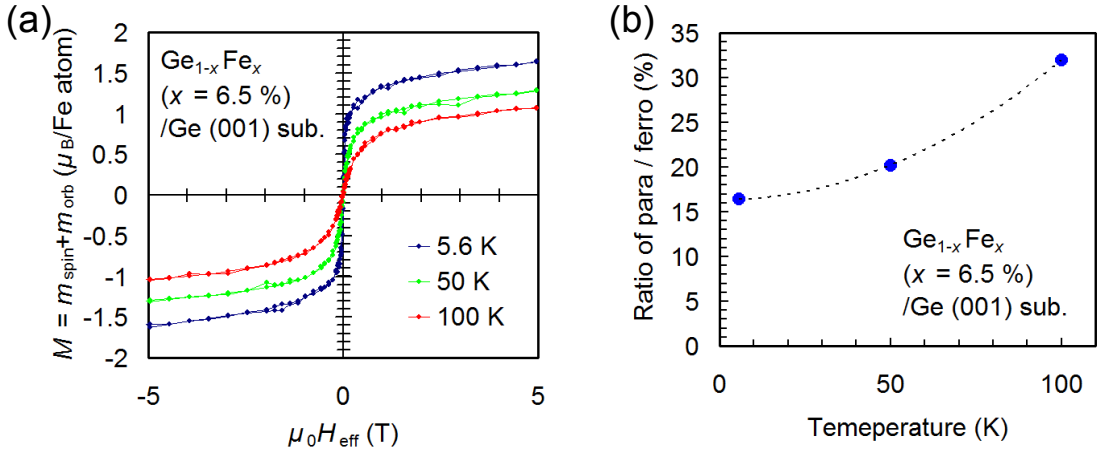

FIG. S3. (a) Effective magnetic field dependence of the total magnetic moment of the $Ge_{1-x}Fe_x$ ($x$ = 6.5 %) film grown on Ge (001) at 240°C (Curie temperature $T_C$ = 100 K) obtained from XMCD study[13]. (b) Temperature dependence of the ratio of the paramagnetic contribution to the ferromagnetic contribution in the total magnetization (ratio of paramagnetic contribution / ferromagnetic contribution) at various temperatures.